\documentclass{article}
\usepackage{graphicx}
\usepackage{epsfig}
\usepackage{amsfonts}
\usepackage{amssymb}
\usepackage{epsf}
\usepackage{cite}
 \usepackage{amssymb}
\usepackage{enumitem}  
\usepackage[dvipsnames]{xcolor}
\usepackage{color} 

\date{\today}

\newcommand{\insertplot}[5]{\begin{figure}
 \hfill\hbox to 0.05in{\vbox to #5in{\vfill
 \inputplot{#1}{#4}{#5}}\hfill}
 \hfill\vspace{-.1in}
 \caption{#2}\label{#3}
 \end{figure}}
 \newcommand{\inputplot}[3]{
 \special{ps: plotfile #1}
}
\newcounter{fig}

\newcommand{\ee}{\end{equation}}
\newcommand{\eea}{\end{eqnarray}}

\begin{document}
\begin{center}

{\Large \bf  Aspects of Gauss-Bonnet scalarisation of charged black holes
}
\vspace{0.8cm}
\\
{{\bf
Carlos A. R. Herdeiro, Alexandre M. Pombo and
Eugen Radu
}
\vspace{0.3cm}
\\
$^{\ddagger }${\small Departamento de Matem\'atica da Universidade de Aveiro and } \\ {\small  Centre for Research and Development  in Mathematics and Applications (CIDMA),} \\ {\small    Campus de Santiago, 3810-183 Aveiro, Portugal}
}
\vspace{0.3cm}
\end{center}

\begin{abstract}
The general relativity vacuum black holes (BHs) can be scalarised in models where a scalar field non-minimally couples to the Gauss-Bonnet (GB) invariant. Such GB scalarisation comes in two flavours, depending on the GB sign that triggers the phenomenon. 
Hereafter these two cases are termed GB$^\pm$ scalarisation. 
For vacuum BHs, only GB$^+$ scalarisation is possible in the static case,
while GB$^-$ scalarisation is $spin$ induced. 
But for electrovacuum BHs, GB$^-$ is also $charged$ induced.
 We discuss the GB$^-$ scalarisation of Reissner-Nordstr\"om and Kerr-Newman BHs, discussing zero modes and constructing fully non-linear solutions. Some comparisons with GB$^+$ scalarisation are given. To assess the generality of the observed features, we also briefly consider the GB$^\pm$ scalarisation of stringy dilatonic BHs and coloured BHs which provide qualitative differences with respect to the electrovacuum case. 
\end{abstract}

 \tableofcontents

\section{Introduction}
 
It is conceivable that deviations from General Relativity (GR) occur \textit{only} for sufficiently large curvatures. One explicit realization of this idea is the phenomenon of spontaneous scalarisation. The original idea proposed the scalarisation of neutron stars in scalar-tensor models~\cite{Damour:1993hw}. More recently, it gained a new guise in which the scalarisation of the GR vacuum BH solutions becomes possible, in the context of extended scalar-tensor models that include the Gauss-Bonnet (GB) quadratic curvature invariant $R^2_{\rm GB} $, as first pointed out in~\cite{Silva:2017uqg,Doneva:2017bvd,Antoniou:2017acq}. In the following we shall dub the latter \textit{GB scalarisation}.

GB scalarisation circumvents well-known no-hair theorems (see~\cite{Herdeiro:2015waa} for a review) due to a certain class of non-minimal couplings 
between a real  scalar field $\phi$ and the GB invariant. The phenomenon occurs for BHs in an appropriate mass range, defined by a dimensionful coupling in  the model. Moreover, it can be triggered either if $R^2_{\rm GB}>0$ - hereafter dubbed GB$^+$ scalarisation~\cite{Blazquez-Salcedo:2018jnn,Myung:2018vug,Doneva:2018rou,Witek:2018dmd,Brihaye:2018bgc,Myung:2018jvi,Minamitsuji:2018xde,Silva:2018qhn,
Brihaye:2018grv,Herdeiro:2019yjy,Hod:2019vut,Collodel:2019kkx,Blazquez-Salcedo:2020rhf,Antoniou:2021zoy,East:2021bqk,Annulli:2021lmn,Doneva:2021tvn} -  or if $R^2_{\rm GB}<0$ - hereafter dubbed GB$^-$ scalarisation. For the Kerr family of GR, the latter only occurs for sufficiently fast spinning BHs~\cite{Dima:2020yac,Herdeiro:2020wei,Berti:2020kgk}, which justifies the terminology \textit{spin-induced scalarisation}~\cite{Dima:2020yac}. By contrast, in the case of GB$^+$ scalarisation, Kerr BHs can also scalarise, but rotation actually suppresses the effects of scalarisation~\cite{Cunha:2019dwb,Collodel:2019kkx}.

\begin{figure}[t!]
\hbox to\linewidth{\hss%
 	\resizebox{9cm}{7cm}{\includegraphics{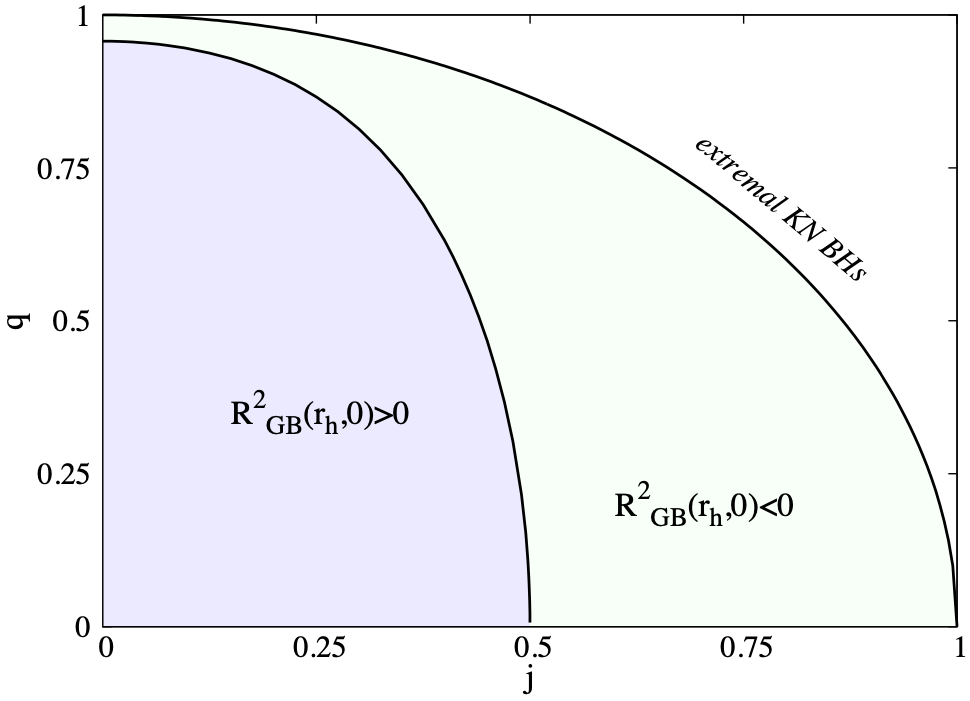}}  
\hss}
\caption{\small  GB invariant  (in units of mass)  of a Kerr-Newman BH with mass $M$, angular momentum $J$ and charge $Q$,
 evaluated at 
the horizon's poles ($r=r_h$ and $\theta=0$), as a function of the dimensionless
parameters $j=J/M^2$ and $q=Q/M$.
Near extremality, $R^2_{GB}<0$.
}
\label{fig1}
\end{figure}

Enlarging the model to include charged BHs, however, GB$^-$ scalarization ceases to rely solely on rotation. This can already be illustrated in electrovacuum GR. The Kerr-Newman solution develops a negative GB invariant for either sufficiently large dimensionless angular momentum $j$ or sufficiently large dimensionless charge $q$. Thus, sufficiently near extremality, Kerr-Newman BHs develop regions with $R^2_{\rm GB}<0$  - Fig.~\ref{fig1}. One may expect that the boundary of the region with $R^2_{\rm GB}<0$ marks the onset of the solutions prone to GB$^-$ scalarisation, as for the Kerr case~\cite{Hod:2020jjy}. We shall confirm this expectation below, explicitly constructing some of the GB$^{-}$ scalarised Kerr-Newman solutions and comparing them with the corresponding GB$^{+}$ scalarised solutions.
 {\small \hspace*{3.cm}{\it  } }

The discussion of the previous paragraph shows that, within electrovacuum BHs, GB$^-$ scalarisation can be spin-induced or charge-induced (or both).
Let us remark, however, that such GB charge-induced scalarisation is different from the scalarisation of charged BHs introduced in~\cite{Herdeiro:2018wub}, where the non-minimal coupling occurs between the scalar field and the Maxwell field, with the GB term (or any curvature corrections) being absent.  

Additionally, the aforementioned observations on the GB$^-$ scalarisation of Kerr-Newman BHs show this process occurs even for spherically symmetric, non-spinning Reissner-Nordstr\"om (RN) BHs~\cite{Brihaye:2019kvj}. Moreover, the negative GB invariant always occurs in the immediate vicinity of the horizon. One may ask whether these features are generic. Is any charged BH model prone to GB$^-$ scalarisation sufficiently close to the maximal allowed charge? And is the $R^2_{\rm GB}<0$  region supporting the scalarisation always occuring in the immediate vicinity of the horizon?  Interestingly, neither of these features is generic, as we shall illustrate by considering two alternative models of charged (spherical) BHs.

This paper is organized as follows.  In Section~\ref{sec2} we detail the Einstein-Maxwell-scalar-GB model and its equations of motion. A general discussion on the tachyonic instability associated with the scalarisation is given, followed by the relevant physical quantities for describing scalarised BHs.  In Section~\ref{sec3} we consider the GB$^\pm$ scalarisation of the RN BH, first discussing the sign of the GB invariant for the electrovacuum RN BH and then constructing both the linear scalar clouds and non-linear scalarised BHs.  In Section~\ref{sec4} we consider the GB$^\pm$ scalarisation of the Kerr-Newman BH. After a brief discussion on the sign of the GB invariant, we discuss the construction of the solutions and provide a sample of numerical results, focusing on the GB$^-$ scalarisation case. In Section~\ref{sec5} we briefly consider Einstein-Maxwell-dilaton and Einstein-non-Abelian BHs, which yield two valuable lessons concerning GB$^\pm$ scalarisation of charged BHs. We conclude with a brief discussion and final remarks in Section~\ref{sec6}.

\section{The Einstein-Maxwell-scalar-GB model}
\label{sec2}
We wish to consider the Einstein-Maxwell-scalar-GB (EMsGB) model, described by the following action
\begin{eqnarray}
\label{action}
\mathcal{S}_{\rm EMsGB}= \int d^4 x \sqrt{-g} 
\left[ \frac{1}{4}R -\frac{1}{4}F_{\mu \nu}F^{\mu \nu}-
 \frac{1}{2}\partial_\mu\phi\partial^\mu\phi 
+\epsilon \frac{\lambda^2 }{4}  f(\phi) R^2_{\rm GB} 
%
\right] \ ,
\end{eqnarray}
where $R$ is the Ricci scalar with respect to the spacetime metric $g_{\mu\nu}$, $R^2_{\rm GB}$  is the GB invariant
\begin{equation}
R^2_{\rm GB} \equiv R_{\alpha\beta\mu\nu} R^{\alpha\beta\mu\nu}-4  R_{\alpha\beta} R^{\alpha\beta}+4R^2 \ ,
\end{equation}  
$R_{\alpha\beta\mu\nu}$ is the Riemann tensor, $R_{\alpha\beta}$ is the Ricci tensor, $F_{\mu \nu}=\partial_\mu A_\nu-\partial_\nu A_\mu$ is the Maxwell field strength tensor where $A=A_\mu  dx^\mu $ is the  $U(1)$ gauge potential, $f(\phi) $ 
is
a coupling function of the real scalar field $\phi$,
$\lambda $ is a constant of the theory
with dimension of length and $\epsilon=\pm 1$ is chosen for GB$^\epsilon$ scalarisation.

Varying the action~(\ref{action}) with respect to the metric tensor
 $g_{\mu \nu}$, 
we obtain 
the Einstein equations, 
\begin{eqnarray}
\label{EGB-eq}
R_{\mu \nu}-\frac{1}{2}g_{\mu \nu} R =2~ T_{\mu \nu}^{\rm (eff)}  \ .
\end{eqnarray}
The effective energy-momentum tensor  $T_{\mu \nu}^{\rm (eff)} $
 has three distinct components:
 \begin{eqnarray}
\label{Teff}
 T_{\mu\nu}^{\rm (eff)}= T_{\mu\nu}^{\rm (s)}+ T_{ \mu\nu}^{\rm (M)}+ T_{ \mu\nu}^{\rm (GB)} \ ,
\end{eqnarray}
consisting of  the (pure) scalar and Maxwell parts, respectively,  
\begin{eqnarray}
T_{\mu \nu }^{\rm (s)}= \partial_\mu \phi \partial_\nu \phi -\frac{1}{2} g_{\mu\nu} \partial_\alpha \phi\partial^\alpha \phi \ ,
~~~~~
T_{\mu \nu }^{\rm (M)}= F_{\mu \alpha}F_{\nu}^{\ \alpha}-\frac{1}{4}F_{\alpha\beta}F^{\alpha\beta} \ ,
 \end{eqnarray} 
and a third contribution  
due to the
 scalar-GB  term in (\ref{action})
 \begin{eqnarray}
\label{Teff2}
 T_{\mu\nu}^{\rm (GB)}=  - 2\epsilon \lambda^2 P_{\mu\gamma \nu \alpha}\nabla^\alpha \nabla^\gamma f(\phi) \ ,
 \end{eqnarray}  
where
 \begin{eqnarray}
\nonumber
		P_{\alpha\beta\mu\nu}  = -\frac14 \varepsilon_{\alpha\beta\rho\sigma} R^{\rho\sigma\gamma\delta} \varepsilon_{\mu\nu\gamma\delta} 
		 = R_{\alpha\beta\mu\nu}+ g_{\alpha\nu} R_{\beta\mu} - g_{\alpha\mu} R_{\beta\nu} + g_{\beta\mu} R_{\alpha\nu}-g_{\beta\nu} R_{\alpha\mu} 
			+\frac12 \left( g_{\alpha\mu}g_{\beta\nu} - g_{\alpha\nu}g_{\beta\mu}\right) R \ ,
 \end{eqnarray}  
and $ \varepsilon_{\alpha\beta\rho\sigma}$ is the Levi-Civita tensor. 
The  scalar field equation is 
\begin{eqnarray}
\label{KG-eq}
\nabla^2 \phi +\epsilon\frac{\lambda^2}{4} \frac{df(\phi)}{d\phi} R^2_{\rm GB}=0 \ ,
\end{eqnarray} 
while the (source-free)
Maxwell equations have the usual form
\begin{eqnarray}
\label{M-eq}
 \nabla_\mu F^{\mu \nu}=0 \ .
\end{eqnarray}

\subsection{GB$^\epsilon$ scalarisation of electrovacuum solutions}
Spontaneous scalarisation manifests itself at the linear level as a tachyonic instability.
Let us assume
 that $\phi=0$ solves~(\ref{KG-eq}), which will hold for a class of coupling functions.  Then, the field equations reduce to those of electrovacuum GR, and the corresponding solutions provide solutions of the full model~(\ref{action}) as well. In particular, the Kerr-Newman geometry will be a solution of this model. Thus, for concreteness we shall refer to the scalarisation of the Kerr-Newman solution  in the following; but a similar discussion would hold for any GR electrovacuum solution.

Next, we consider scalar perturbations of the Kerr-Newman solution within the full model~(\ref{action}). 
Assuming a small-$\phi$
expansion for the coupling function
\begin{eqnarray}
\label{f-phi-small}
f(\phi)=f\big|_{\phi=0}+ \frac{1}{2} \frac{d^2 f}{d \phi^2}\Big|_{\phi=0}  \phi^2+\mathcal{O}(\phi^3) \ ,
\end{eqnarray}
the linearised scalar field equation~(\ref{KG-eq}) around the Kerr-Newman solution becomes
\begin{eqnarray}
\label{eq-phi-small}
(\Box-\mu_{\rm eff}^2)\phi =0 \ , \qquad {\rm where} \qquad  \mu_{\rm eff}^2=  
-\epsilon\frac{ \lambda^2}{4}  \frac{d^2 f}{d \phi^2}\Big |_{\phi=0}R^2_{\rm GB} \  , 
\end{eqnarray} 
where $\Box$ and $R^2_{\rm GB}$ are computed for the scalar-free Kerr-Newman solution.

If $\mu_{\rm eff}^2$ is not strictly positive, the scalar field possesses 
a (spacetime dependent) tachyonic mass. Wherever this tachyonic mass is supported, such region potentially supports a spacetime instability, which is precisely the GB scalarisation.  
To simplify the discussion,  we assume 
without any loss of generality that  
${d^2 f}/{d \phi^2}\Big |_{\phi=0} $
is strictly positive.
Then the condition
$\mu_{\rm eff}^2<0$
is equivalent to
\begin{eqnarray}
\label{conds}
\epsilon R^2_{\rm GB}>0~.
\end{eqnarray}
When this condition is obeyed in some region(s) outside the 
BH horizon, GB$^\epsilon$ scalarisation is triggered.

If the Kerr-Newman BH reduces to a Schwarzschild BH  of mass $M$, then
\begin{eqnarray}
\mu_{\rm eff}^2=  - \epsilon \frac{\lambda^2}{4}\frac{d^2 f}{d \phi^2}\Big |_{\phi=0}  \frac{48 M^2}{r^6} \ , 
\end{eqnarray}
and only GB$^{+}$ scalarisation is possible.

\subsection{Physical quantities of interest for scalarised BHs}
When the above instability is present there is also a different class of solutions for the model~(\ref{action}), besides the electrovacuum ones. These are the scalarised solutions. We are interested in the case of stationary BHs. 
Generically, these solutions posses three global charges:
the mass $M$, the electric charge $Q$
and the angular momentum $J$.
For the solutions in this work, there is also a ``scalar charge" $Q_s$,
which is not, however, associated  with a conservation law.
There are also a number of relevant horizon quantities: the Hawking temperature $T_H$,
the horizon area $A_H$, the entropy $S$, and
the  horizon angular velocity $\Omega_H$.
The BH entropy is the sum of two terms,
\begin{eqnarray}
\label{S}
S=S_{\rm E}+S_{\rm sGB}\ ,{~~}{\rm with} \qquad 
S_{\rm E}=\frac{1}{4}A_H\ , \qquad  S_{\rm sGB}= \epsilon \frac{1}{2}\lambda^2 \int_{H} d^2 x \sqrt{h}f(\phi) {\rm  R}^{(2)} \ ,
\end{eqnarray}
where
${\rm  R}^{(2)}$ is the Ricci scalar of the induced horizon metric $h$.
The solutions satisfy the  Smarr law
\begin{eqnarray}
\label{smarr}
M=2\Omega_H J+2 T_H S+\Phi Q+ M_s \ ,
\end{eqnarray}
where $\Phi$  is the electrostatic potential 
and
 $M_s$ is a contribution of the scalar field
\begin{eqnarray}
\label{sup}
 M_s= \frac{1}{2} \int d^3g \sqrt{-g} (\partial_a \phi)^2~.
\end{eqnarray}
Also,
the solutions satisfy 
the first law of BH thermodynamics
\begin{eqnarray}
\label{first-law} 
dM=T_H dS +\Omega_H dJ +\Phi dQ \ ,
\end{eqnarray}
in which there is no contribution from the scalar field.

In this work we shall focus on the quadratic coupling function,\footnote{For spherical symmetry, we have also explored the exponential coupling studied in \cite{Doneva:2018rou,Doneva:2017bvd}, and observed that the behaviour is qualitatively similar.}
\begin{eqnarray}  
\label{fu}
f(\phi)=\frac{\phi^2}{2} \ ,
\end{eqnarray}  
which is the simplest choice of $f(\phi)$ that guarantees that $\phi=0$ satisfies the scalar equation~(\ref{KG-eq}).

It is useful to observe that
the equations of the model are invariant under the transformation
\begin{eqnarray}
\label{scale}
r\to \alpha r \ , \qquad \lambda  \to \alpha \lambda \ ,
\end{eqnarray} 
 with 
$r$ the radial coordinate and
$\alpha>0$ an arbitrary  positive constant.
Only quantities invariant under (\ref{scale})
($e.g.$ $Q/M$ or $Q/\lambda$)
have a physical meaning.
Following standard terminology, we define the 
following
\textit{reduced} quantities:
\begin{eqnarray}
\label{scale1}
q\equiv \frac{Q}{M}\ , \qquad a_H\equiv \frac{A_H}{16\pi M^2}\ , \qquad t_H\equiv 8\pi T_H M \ , \qquad j\equiv \frac{J}{M^2} \ ,
\end{eqnarray}
which will be considered in what follows.

\section{GB$^\epsilon$ scalarisation of Reissner-Nordstr\"om BHs}
\label{sec3}
Let us start by considering the spinless limit of the Kerr-Newman family, the RN BH.
The corresponding metric and gauge field can be written as (see $e.g.$~\cite{Townsend:1997ku})
\begin{eqnarray}  
\label{RNn} 
ds^2=-N(r) \sigma(r)^2  dt^2+\frac{dr^2}{N(r)}+r^2(d\theta^2+\sin^2 \theta d\varphi^2) \ , \qquad V=V(r)dt \ ,
\end{eqnarray}
where 
\begin{eqnarray}
\sigma(r)=1 \ , \qquad  N(r)=1-\frac{2M}{r}+\frac{Q^2}{r^2}\ , \qquad  V(r)=\frac{Q}{r} \ .
\label{RNn2} 
\end{eqnarray}
This BH possess an 
event horizon at
\begin{eqnarray}   
r=r_h=M+\sqrt{M^2-Q^2} \ .
\label{hr}
\end{eqnarray}
Thus, $0\leqslant q\leqslant 1$ and $q=1$ for the extremal RN BH.

The GB invariant of the RN metric reads
\begin{eqnarray}   
R^2_{\rm GB}   = \frac{8}{r^8}\left[6M^2r^2-12Q^2Mr
+
5 Q^4
\right] \ .
\label{gbrn}
\end{eqnarray}
For $Q\neq 0$ this always becomes negative for some region with $r>0$.
 This region, however, is cloaked by the 
horizon unless the largest root of the quadratic equation in (the square brackets in)~(\ref{gbrn}) exceeds $r_h$. This condition is
\begin{equation}
qQ\left(1+\frac{1}{\sqrt{6}}\right)>r_h \ .
\end{equation}
Using~(\ref{hr}) one can easily show that is possible for
\begin{eqnarray}   
q>q_{c}\simeq 0.957058 \ . 
\end{eqnarray}
Thus, for $q_c<q\leqslant 1$, a RN BH can undergo GB$^-$ scalarisation.

\subsection{The linear scalar clouds}
At the onset of the tachyonic instability, the linearised scalar field equation (\ref{eq-phi-small}) on the RN background allows solutions known as \textit{scalar clouds}. These occur for a discrete set of RN solutions, each solution corresponding to a particular harmonic scalar field mode. To see this, we perform a harmonic decomposition of the scalar field,
\begin{eqnarray}
\label{aqw}
\phi=U(r) Y_{\ell m}(\theta,\varphi) \ ,
\end{eqnarray}
where  $Y_{\ell m}$ are the real spherical harmonics and $\ell,m$ are
the associated quantum numbers with the usual ranges, $\ell=0,1,\dots$ and $-\ell \leqslant m \leqslant \ell$.
For a RN BH background (\ref{RNn})-(\ref{RNn2}), the linearised scalar equation (\ref{eq-phi-small}) becomes a radial equation
\begin{eqnarray}
\label{aqw1}
\left[r^2 N(r)U'(r)\right]'=\ell(\ell+1)U(r)-\epsilon \frac{\lambda^2}{r^2} 
\left(
\frac{12 M^2}{r^2}+\frac{10Q^4}{r^4}-\frac{24 MQ^2}{r^3}
\right)~,
\end{eqnarray}
where a prime denotes a derivative $w.r.t.$ the radial coordinate $r$.

This equation has the following asymptotic solutions: near the horizon
\begin{eqnarray}
\label{sph1}
U(r)=u_0+\frac{r_h}{Q^2-r_h^2}
\left[
-  \ell(\ell+1)+\epsilon \frac{\lambda^2}{r_h^2} \left(3-\frac{6Q^2}{r_h^2}+\frac{Q^4}{r_h^4}\right)
\right]u_0
(r-r_h)+\dots \ , 
\end{eqnarray}
where $u_0$ is an arbitrary nonzero constant which, in numerics, we set to $1$; 
and near spatial infinity
\begin{eqnarray}
\label{sph2}
U(r)= \frac{Q_s}{r^{\ell+1} } +  \dots \ ,
\end{eqnarray}
where $Q_s$ is the scalar charge for $\ell=0$.

Solving (\ref{aqw1}) with the above asymptotic behaviours can be viewed as an eigenvalue problem.
For a given $\ell$, requiring the scalar field to smoothly interpolate between the asymptotics 
(\ref{sph1}) and (\ref{sph2}), crossing $n$ times the $r$-axis, 
selects a discrete set of RN parameters. Thus, each scalar cloud is characterized by three 'quantum numbers' $(n, \ell, m)$, 
with the radial function being degenerate in terms of $m$ as a consequence of the spherical symmetry of the RN background. In this work we shall report results on  nodeless spherically symmetric fundamental solutions only\footnote{Similar solutions are likely to exist for any other values of  the quantum numbers,
some preliminary results being found for the $\ell=1$, $m=n=0$ case.
An investigation of $Q=0$, 
$\ell=1$ static solutions has been reported in~\cite{Collodel:2019kkx}.}, $i.e.$ with $\ell=m=n=0$.

\begin{figure}[t!]
\hbox to\linewidth{\hss%
   \resizebox{9cm}{7cm}{\includegraphics{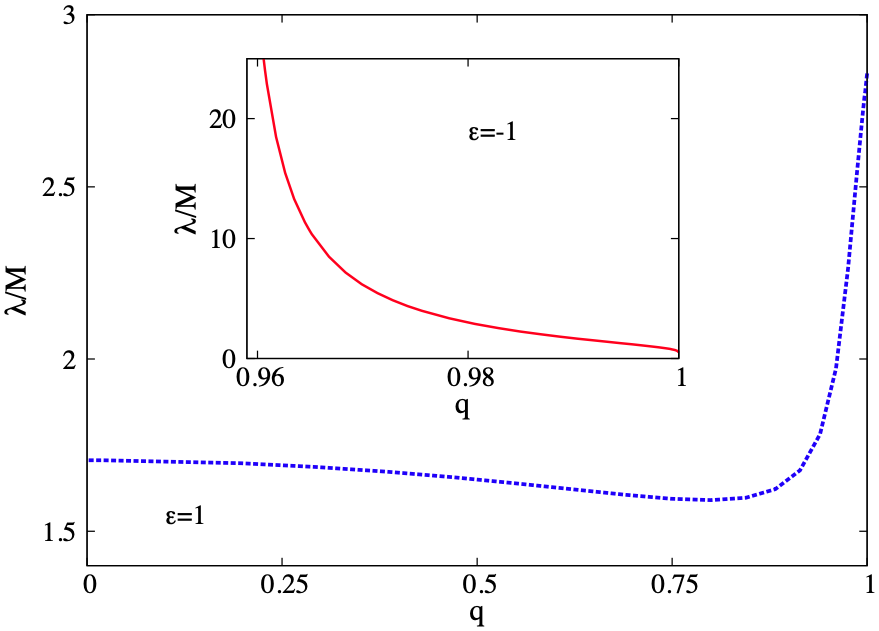}}  
	\resizebox{9cm}{7cm}{\includegraphics{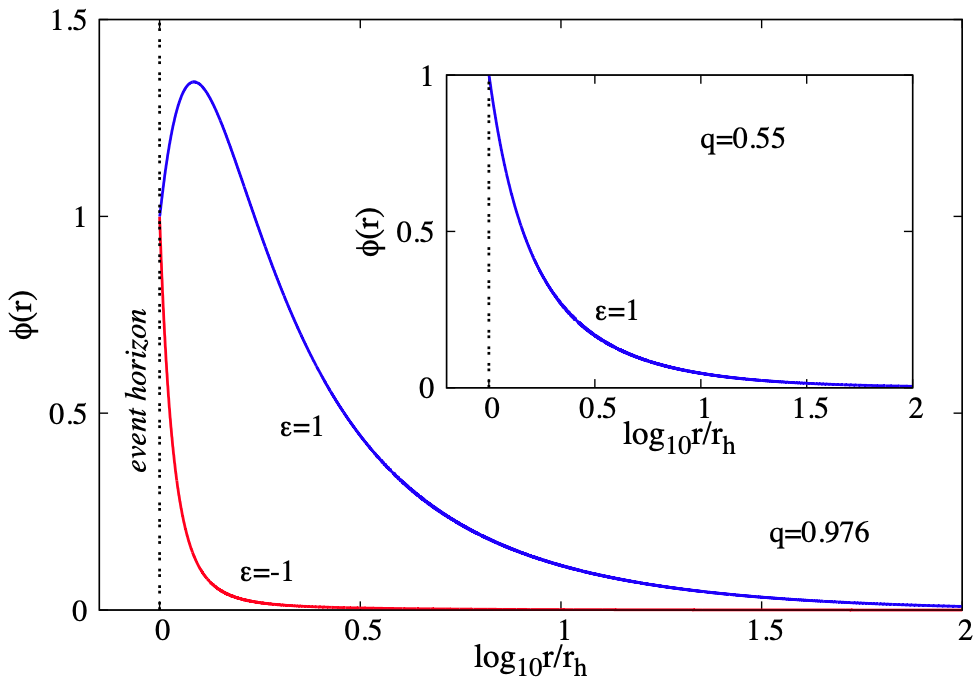}} 
\hss}
\caption{\small 
{\it Left panel:}  Dimensionless ratio $\lambda/M$ of the set of  RN  solutions supporting the $\ell=m=n=0$ scalar cloud $vs.$ the reduced electric charge $q$ for $\epsilon=\pm 1$.
{\it Right panel:} Typical radial profiles of the spherical, nodeless scalar clouds on a RN BH background. 
}
\label{figRN1}
\end{figure}

Having chosen the clouds quantum numbers, taking $\lambda$ as fixed scale set in the action, and fixing the reduced charged $q$, the radial equation has a solution for a specific dimensionless ratio $\lambda/M$. For instance, for $\epsilon=+1$,  $\ell=m=n=0$ and $q=0$ the selected value is $\lambda/M\sim 1.704$, corresponding to the initial point of the blue dotted curve in  Fig.~\ref{figRN1} (left panel). This is the zero mode of the GB$^+$ instability of Schwarzschild. It selects a mass scale. Smaller masses (larger $\lambda/M$) describe BHs unstable against scalarisation; larger masses  (smaller $\lambda/M$) correspond to stable BHs. 
 {\small \hspace*{3.cm}{\it  } }

The variation of $\lambda/M$ with increasing $q$ can be interpreted as follows. GB scalarisation of Schwarzschild BHs may be attributed to a repulsive gravitational effect of the GB term, which only becomes dominant for sufficiently small BHs (in terms of $\lambda$). Adding electric charge introduces two competing effects. On the one hand the electric charge provides a repulsive gravitational effect for RN BHs. This facilitates scalarisation, making it available for larger BHs (larger $M$, smaller $\lambda/M$). On the other hand, the repulsive gravitational effect of the GB invariant, which is at the source of the scalarisation phenomenon, becomes supressed (and eventually the GB term even changes sign in some region), when increasing $q$. This supresses scalarisation, making it available only for smaller BHs (smaller $M$, larger $\lambda/M$). The trend observed in the blue dotted curve in the main panel of  Fig.~\ref{figRN1} (left panel) suggest that for small $q$ the RN charge repulsion dominates and for large $q$ the GB charge suppression becomes dominant.  The inset, on the other hand, gives the behaviour for $\epsilon=-1$. In this case, as $q$ increases, in the allowed (large $q$) interval, the GB behaviour dominates, and due to the opposite sign it  provides an ever more significant repulsive contribution, thus faciliating GB$^-$ scalarisation, which therefore is available for larger masses. 
We also remark that,  for GB$^-$ scalarisation, the ratio $\lambda/M$ appears to diverge as $q \to q_{c}$, 
while it stays finite as $q\to 1$.

The  profiles of typical scalar clouds are shown in Fig.~\ref{figRN1} (right panel).
For $\epsilon=+1$ and moderate $q$ we recover the picture found in the Schwarzschild case, $Q=0$: a monotonically decreasing profile starting with some finite value at the horizon (see the inset). This is also true for the $\epsilon=-1$ case (red curve in main panel).
But for  $q>q_{c}$ and $\epsilon=+1$, a new qualitative behaviour emerges:  the maximal value of the scalar cloud can be attained outside the horizon\footnote{ 
This feature can be explained  by studying eq. (\ref{aqw1})~\cite{Brihaye:2019kvj}.
} - see the blue curve in the main panel.

\subsection{The non-linear spherically symmetric scalarised BHs}
The linear scalar clouds just discussed can be continued to the non-linear regime. Their backreaction originates scalarised BHs. 
We shall now discuss their construction for the case of the spherical, nodeless scalar clouds.

The ansatz to obtain the scalarised BH solutions is~(\ref{RNn}), together with a radial scalar field
\begin{eqnarray}
\phi\equiv \phi(r) \ .
\end{eqnarray}
As in the usual electrovacuum case,
	the Maxwell equation~(\ref{M-eq}) yields a total derivative,
leading to a first integral 
\begin{eqnarray}
V'(r)=-\frac{Q\sigma(r)}{r^2} \ .
\end{eqnarray}
This introduces the electric charge measured at infinity, $Q$.
The scalar field satisfies the equation
\begin{eqnarray}
\phi''
+\left(
\frac{2}{r}+\frac{N'}{N}+\frac{\sigma'}{\sigma}
\right)
\phi'
- \frac{\epsilon \lambda^2}{r^2 N\sigma}
\Big[
(3-5N)N'\sigma'
+\sigma \big( (1-N)N''-N'^2)
+2(1-N)N\sigma''
\Big] \phi=0~,
\end{eqnarray}

while the equations 
for the metric functions $N,\sigma$
are too involved and shall not be displayed here.

We are interested in BH solutions with an event horizon  located at $r=r_h>0$ . 
The equations of the model are subject to
  the following boundary conditions.
\begin{equation}
N\big|_{r_h}=0\ , \quad \sigma\big|_{r_h}=\sigma_h\ , \quad \phi\big|_{r_h}=\phi_h\ ,\quad V\big|_{r_h}=0 \ ;
\quad N\big|_{\infty}=1\ , \quad \sigma\big|_{\infty}=1 \ ,  \quad \phi\big|_{\infty}=0\ , \quad V\big|_{\infty}=\Phi \ ,
\end{equation}
where $\sigma_h$, $\phi_h$ are constants fixed by numerics
and $\Phi$ is the electrostatic potential at infinity.
The horizon data fix the  Hawking temperature,
the horizon area  and the entropy of the solutions,
\begin{eqnarray}
T_H=\frac{\sigma_h N'(r_h)}{4\pi}\ , \qquad A_H=4\pi r_h^2 \ , \qquad S=\pi r_h^2+\epsilon \lambda^2 f(\phi_h) \ .
\end{eqnarray}
A local solution compatible with these asymptotics can be constructed both
close to the horizon (as a power series in $(r-r_h)$) and at infinity  (as a power series in $1/r$).
For example, the first terms in the far field expression of the solutions  read
\begin{equation}
\label{inf1}
N=1-\frac{2M}{r}+\frac{Q^2+Q_s^2}{r^2}+\dots \ , \quad \phi(r)=\frac{Q_s}{r}+\dots \ , \quad 
V(r)=\Phi-\frac{Q}{r}+\dots\ , \quad 
\sigma(r)=1-\frac{Q_s^2}{2r^2}+\dots \ .
\end{equation}

 With the details just laid out we have numerically constructed the nonlinear continuation of 
the scalar clouds solving the full equations of the EMsGB model, for both signs of $\epsilon$. 

\begin{figure}[t!]
\hbox to\linewidth{\hss%
	\resizebox{9cm}{7cm}{\includegraphics{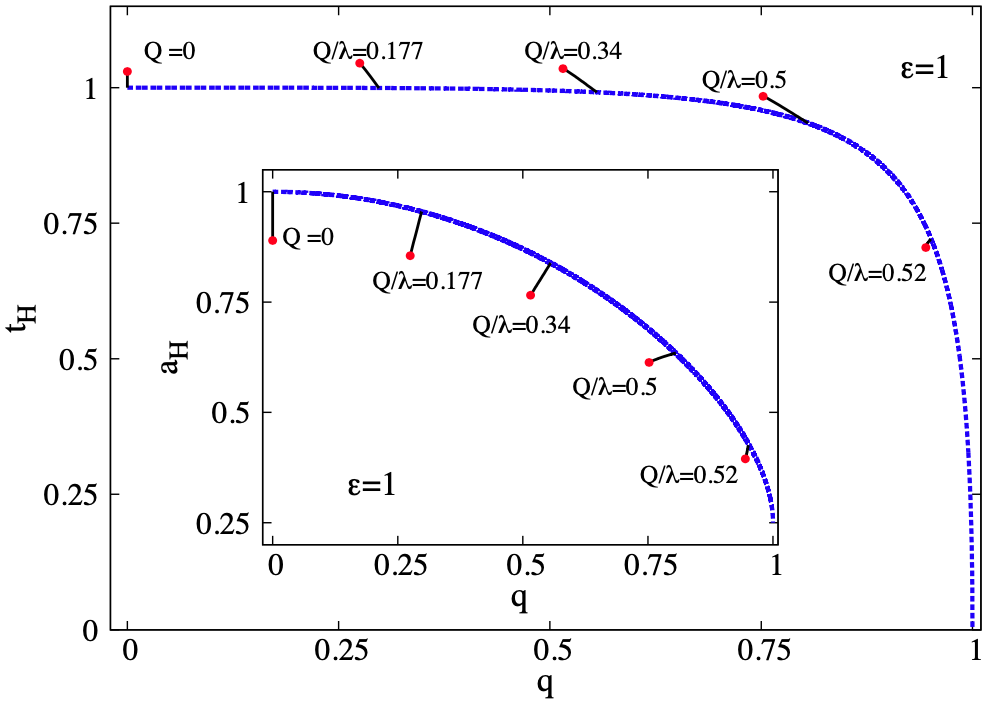}} 
 	\resizebox{9cm}{7cm}{\includegraphics{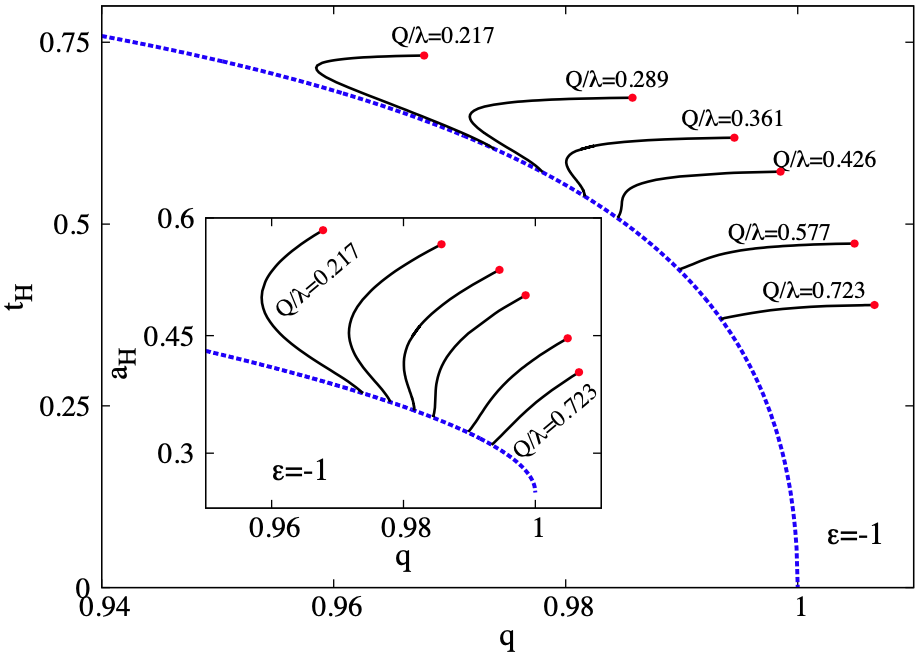}}  
\hss}
\caption{\small 
 Charge-temperature (main panels panel) and charge-horizon area (insets)
diagrams, in units set by mass,  for several illustrative families
of GB$^\epsilon$ scalarised RN solutions, for both $\epsilon= +1$ (left panel) and $\epsilon= -1$ (right panel).
The branches of scalarised solutions bifurcate from the
electrovacuum RN BHs (blue line) and terminate in critical configurations (red circles).
}
\label{figRN2}
\end{figure}

Technically, the construction of the scalarised BHs is a one parameter shooting problem
in terms of the value of the scalar field at the horizon $\phi_h$.
The input parameters are $r_h,Q$ and $\lambda$.
Fixing the length scale $\lambda$, this leads to
a two dimensional parameter
 space for the problem.
The  numerical results for several values of the ratio $Q/\lambda$
are shown in Fig.~\ref{figRN2}.
 {\small \hspace*{3.cm}{\it  } }

Fig.~\ref{figRN2} shows that for a given ratio $Q/\lambda$ and both values of $\epsilon$, one finds a continuum of  solutions which bifurcate from the corresponding
 RN BH 
supporting a scalar cloud with these parameters.
This line has a finite extent, 
ending in a critical configuration where
the
numerical process fails to converge.
A general explanation for this behaviour can be traced back to the
fact that the radicand of a square root in the horizon
expansion of the scalar field vanishes as the critical set
is approached. This is a generic feature of GB-scalar models. 
An exception here are the
$\epsilon=+1$
 solutions emerging from RN BHs with $q>q_c$,
in which case the critical configurations seem
to possess a curvature singularity for some radius outside the event horizon (see~\cite{Brihaye:2019kvj} for a discussion).
 
From Fig.~\ref{figRN2} one may highlight two qualitatively different features when comparing $\epsilon=\pm 1$.
First, scalarisation reduces (increases) $a_H$  for $\epsilon=+1$ ($\epsilon=-1$). Secondly, ``overcharged" solutions with $q>1$
 exist for $\epsilon=-1$
only.

\section{GB$^\epsilon$ scalarisation of Kerr-Newman BHs}
\label{sec4}

Let us now address the GB$^\epsilon$ scalarisation of the spinning, charged electrovacuum Kerr-Newman BH. This is a solution of the model (\ref{action}), with the coupling (\ref{fu}), together with 
a vanishing scalar field,
$\phi=0$.
This BH is described by its ADM mass $M$, total angular momentum per unit mass $a=J/M$ and electric charge $Q$.
In Boyer-Lindquist coordinates it reads (see $e.g.$~\cite{Townsend:1997ku})
\begin{equation}
\label{KerrNewmanMetric}
ds^2 = - \frac{\Delta}{\Sigma} \left( dt -a \sin^2 \theta d\varphi \right)^2 +  \Sigma \left( \frac{dr^2}{\Delta} + d\theta^2 \right) + \frac{\sin^2 \theta}{\Sigma} \left[ a dt - \left( \Sigma+a^2\sin^2\theta \right) d\varphi \right]^2 \ ,
\end{equation}
and
\begin{equation}
\label{ElectromagneticVectorKerrNewman}
A_\mu dx^\mu = -\frac{Q r}{\Sigma} \left( dt - a \sin^2 \theta d\varphi \right) \ ,
\end{equation}
where 
\begin{equation}
\Delta \equiv r^2 - 2Mr + a^2 + Q^2 \ , \qquad  \Sigma \equiv r^2 + a^2 \cos^2 \theta \ .
\label{deltasigma}
\end{equation}
The event horizon of this solution is located at
\begin{equation}
r_h=M \left (1+\sqrt{1-j^2-q^2} \right).
\end{equation}
This implies the Kerr-Newman bound, $j^2+q^2 \leqslant 1$; an extremal BH saturates this bound. 

The Kerr-Newman metric has a GB invariant
\begin{eqnarray}
\label{GBKN}
R^2_{\rm GB}   =&& \frac{48 M^2}{\Sigma^3}
\left(
1-\frac{2a^2}{\Sigma^3}\big(3r^2-a^2 \cos^2\theta\big)^2\cos^2\theta
\right) 
\\
\nonumber
&&
+\frac{8Q^2}{\Sigma^6}
\Big\{
r^4(5Q^2-12Mr)+a^2\cos^2\theta \Big[2r^2(-19Q^2+60 Mr)
+5a^2(Q^2-12Mr)\cos^2\theta\Big] 
\Big\}~.
\end{eqnarray}

\begin{figure}[t!]
\hbox to\linewidth{\hss%
	\resizebox{9cm}{7cm}{\includegraphics{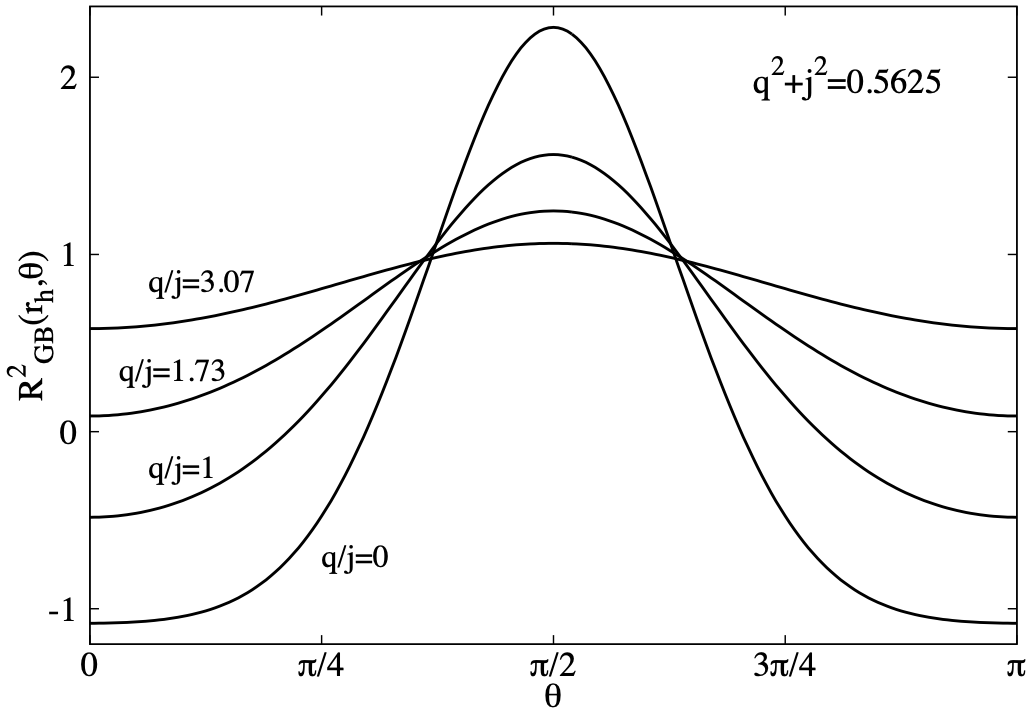}} 
\hss}
\caption{\small 
The GB invariant at the horizon
(in units of mass)
 as a function of the $\theta-$coordinate for several Kerr-Newman BHs.
}
\label{figKN1}
\end{figure}

We have studied\footnote{An alternative expression for (\ref{GBKN}),  in terms of 
$P_1\equiv (1+\sqrt{1-j^2+q^2}){r}/{r_h}$ and 
$P_2\equiv j \cos \theta$, is
\begin{eqnarray}
\nonumber
R^2_{\rm GB}   = 
\frac{48}{M^4 }\frac{1}{(P_1^2+P_2^2)^3}
\bigg[
1-\frac{2}{(P_1^2+P_2^2)^3}
   \bigg(
	P_2^2(3P_1^2-P_2^2)^2
	+q^2  \big(
	P_1(P_1^4-10P_1^2P_2^2+5P_2^4)
~~
	-\frac{q^2}{12} \left(5P_1^4-38 P_1^2P_2^2+5P_2^4 \right)
	      \big)
	 \bigg)
\bigg]~,
\end{eqnarray}
a form
which has been employed in our study.
}
the sign of this quantity as a function of the 
parameters $(j,q)$ -  Fig.~\ref{fig1}, observing  that the qualitative picture
found in the Kerr ($q=0$) or RN (j=0) cases
is still valid for a Kerr-Newman BH.
While $R^2_{\rm GB}$
is positive for large values of the radial coordinate,
its sign close to the event horizon depends on the value of
$(j,q)$.
That is, for fixed $j$ (or $q$),
the GB invariant $R^2_{\rm GB}$
always becomes negative in a region outside the horizon,
for large enough values of  $q$ (or $j$).
In the presence of rotation, this region is located around the poles of the horizon,
$\theta=0,\pi$ - Fig. \ref{figKN1}.
In Fig. \ref{fig1}
we show the region in the $(j,q)$-domain where the 
 GB invariant takes a negative sign at  the poles of the horizon.
Kerr-Newman BHs with $R^2_{\rm GB}<0$
have the potential to be scalarised for both signs of $\epsilon$.

 {\small \hspace*{3.cm}{\it  } }

 \subsection{Construction of the scalarised Kerr-Newman BHs}
To construct the GB$^\epsilon$ scalarised Kerr-Newman BHs, we shall use the ansatz in 
\cite{Cunha:2019dwb,Herdeiro:2020wei}, 
 supplemented with a  nonzero gauge field.
The metric ansatz reads:\footnote{The Kerr-Newman BH can also be written in this coordinate system. The corresponding expressions in the Kerr limit can be found in~\cite{Herdeiro:2015gia}.}
\begin{eqnarray}
\label{ansatz}
ds^2=-e^{2F_0} N dt^2+e^{2F_1}\left(\frac{dr^2}{N }+r^2 d\theta^2\right)+e^{2F_2}r^2 \sin^2\theta (d\varphi-W dt)^2\ , \qquad 
~{\rm with} \quad N=1-\frac{r_h}{r}\ ,
\end{eqnarray} 
where the metric functions
$F_i,W$, as well as the scalar field ansatz $\phi=\phi(r,\theta)$, 
depend on $r,\theta$ only.
The ansatz for the $U(1)$ potential is
\begin{eqnarray}
\label{ansatzA}
 A=A_\varphi \sin \theta (d\varphi - W dt) + V dt \ ,
\end{eqnarray} 
with two potentials, a magnetic one $A_\varphi$ and an electric one $V$, both depending on $r,\theta$.

Setting $A_\varphi=V=0$ results in the spontaneously scalarised Kerr BHs in~\cite{Cunha:2019dwb,Herdeiro:2020wei}, albeit with a different coupling function.
The limit $W=A_\varphi=0$ results in the scalarised RN BHs~\cite{Herdeiro:2018wub,Fernandes:2019rez,Astefanesei:2019pfq}
discussed above, albeit for a different radial coordinate.

The general problem is solved subject to the following boundary conditions.
Asymptotic flatness requires
\begin{equation}
\lim_{r\rightarrow \infty}{F_i}=\lim_{r\rightarrow \infty}{W}=0 \ ,
 \qquad {\rm and} \qquad \lim_{r\rightarrow \infty}{\phi}=\lim_{r\rightarrow \infty}{A_\varphi}=0 \ , \qquad \lim_{r\rightarrow \infty } V=\Phi \ .
\end{equation}
Axial symmetry and regularity impose
the following boundary conditions on the symmetry axis, $i.e.$ at $\theta=0,\pi$:
\begin{equation}
\partial_\theta F_i = \partial_\theta W = \partial_\theta \phi =\partial_\theta V= A_\varphi=0 \ .
\end{equation}
Moreover, the absence of conical singularities implies also that 
%
$
F_1=F_2 
$
%
on the symmetry axis. 
 
The event horizon is located at a constant value of $r=r_h>0$.
Only non-extremal solutions can be studied within the metric ansatz
(\ref{ansatz}). We proceed 
 by introducing a new
radial coordinate 
%
$x=\sqrt{r^2-r_H^2}$, which simplifies  the boundary conditions at the horizon
 and also the numerical treatment of the problem.
This results in the following boundary conditions at the horizon 
\begin{equation}
\partial_x F_i \big|_{r=r_h}= \partial_x \phi  \big|_{r=r_h} =  0 \ , \qquad W \big|_{r=r_h}=\Omega_H \ , \qquad 
\partial_x A_\varphi \big|_{r=r_h}=V \big|_{r=r_h}=0 \ ,
\end{equation}
where the constant $\Omega_H>0$ is the horizon angular velocity.
Then, 
an approximate expansion of the solution compatible with these boundary conditions
can easily be constructed.

Specializing some of the aforementioned physical quantities of interest for the ansatz in use, we obtain that 
the Hawking temperature and the event horizon area 
are determined by the following horizon data,
\begin{eqnarray}
\label{THAH}
&&
T_H=\frac{1}{4\pi r_h}e^{F_0^{(0)}(\theta)-F_1^{(0)}(\theta)} \ ,
\qquad 
A_H=2\pi r_h^2 \int_0^\pi d\theta \sin \theta~e^{F_1^{(0)}(\theta)+F_2^{(0)}(\theta)} \ , 
\end{eqnarray}
with the near horizon expansion $F_i=F_i^{(0)}(\theta)+x^2 F_i^{(2)}(\theta)+\dots$, and 
$i=0,1,2$.

The ADM mass $M$,  the angular momentum $J$,
 the scalar 'charge' $Q_s$,
together with the
  the magnetic dipole momentum $q_m$, the electrostatic potential $\Phi$  and 
	the electric charge $Q$
 are read off from 
the far field asymptotic of the metric and matter functions  
\begin{equation}
\label{asym}
g_{tt} =-1+\frac{2M}{r}+\dots\ ,~~g_{\varphi t} =-\frac{2J}{r}\sin^2\theta+\nonumber \dots\ , 
~~\phi=-\frac{Q_s}{r}+\dots \ ,~~
A_\varphi=\frac{q_m \sin \theta }{r}+\dots \ ,~~V=\Phi-\frac{Q}{r}+\dots \ .
\end{equation}  
We remark that both the metric functions and the scalar field are invariant $w.r.t.$ the transformation $\theta \to \pi-\theta$.

 {\small \hspace*{3.cm}{\it  } }
\begin{figure}[t!]
\hbox to\linewidth{\hss%
	\resizebox{9cm}{7cm}{\includegraphics{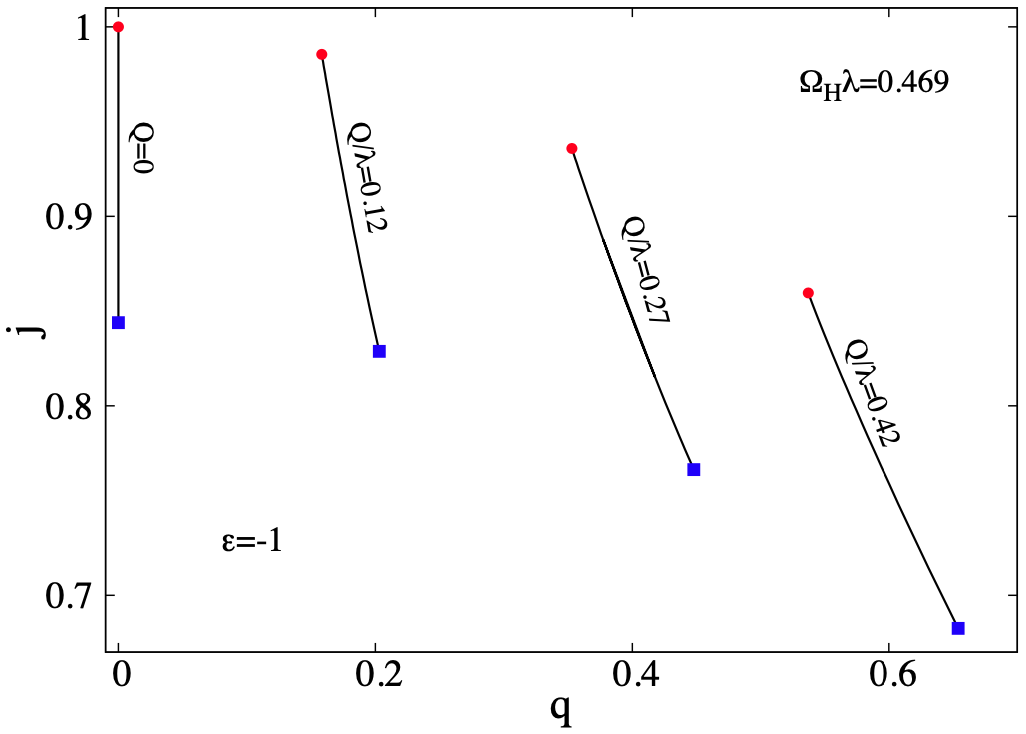}} 
 	\resizebox{9cm}{7cm}{\includegraphics{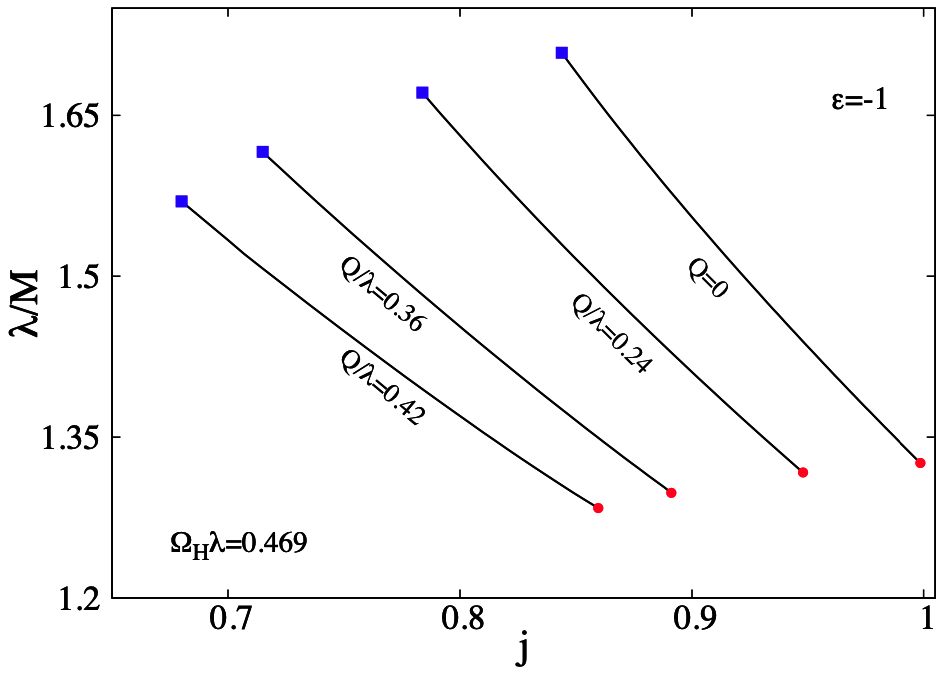}}  
\hss}
\caption{\small 
Branches of GB$^{-}$ scalarised Kerr-Newman BHs in a q $vs.$ $j$  plot (left panel) and $j$ $vs.$ $\lambda/M$ plot (right panel). The branches are for a specific choice of $\Omega_H\lambda=0.469$ and different choices of $Q/\lambda$.
The blue squares correspond to  Kerr-Newman BHs with a vanishing
scalar field, while the red circles correspond to critical configurations.
}
\label{figKN2}
\end{figure}


 \subsection{Numerical results}

With the setup just described, we have employed a numerical approach similar to the one in~\cite{Cunha:2019dwb,Herdeiro:2020wei}.
The typical numerical errors for the solutions so obtained and reported below are  of the order of $10^{-3}$.

The scalarised Kerr-Newman solutions possess four independent charges: the three global charges shared with their  electrovacuum counterparts, $(M,J,Q)$, plus the scalar charge $Q_s$.
In our approach, the input parameters are:
the event horizon radius 
$r_h$, the  horizon angular velocity $\Omega_H$, 
the asymptotic value of the electrostatic potential $\Phi$
(or the electric charge $Q$),
together with the coupling constant $\lambda$ (which specifies  the theory). 
Therefore, after fixing the scale $\lambda$, we are left with a three dimensional parameter space. 
A full scanning of such large parameter space  of scalarised Kerr-Newman BHs is therefore a time consuming task. 
Here we focus on illustrative sets of solutions which already capture the generic behaviour. Moreover, 
although we have verified that spinning scalarised solutions exist for both signs of $\epsilon$, 
we shall focus on the results for the more novel case of spin/charge scalarisation, $\epsilon =-1$.
In practice we have scanned the parameter space by varying both $r_h$ and 
$Q$,
for several different  values of 
$\Omega_H$. 
%
Alternatively, we have varied both $r_h$ and 
$\Omega_H$,
for fixed values of 
$Q$.

Our numerical results suggest that the solutions share most of the properties of the 
scalarised Kerr BHs discussed in~\cite{Herdeiro:2020wei,Berti:2020kgk}.
In Fig.~\ref{figKN2} we display 
the reduced quantities
 $(q,j)$ (left panel) and $(j, M/\lambda)$
(right panel)
 for solutions 
with $\Omega_H \lambda= 0.469$
and illustrative values of the ratio $Q/\lambda$.
These scalarised solutions emerge from a Kerr-Newman BH supporting a zero mode solution of the scalar equation - a scalar cloud - corresponding to the blue squares. Then, the sequence of solutions with constant $Q/\lambda$ terminates at a critical configuration, as in the RN case reported before, corresponding to the red circles. 
The main trends observed in Fig.~\ref{figKN2} is that fixing $Q$ and $\Omega_H$ in units of $\lambda$, the scalarised BHs have more mass (and thus smaller $q$ and $\lambda/M$) and larger $j$.

Although our scanning of the full parameter space was limited, extrapolating the
  existing numerical data, we antecipate that the (three dimensional) domain of existence of $\epsilon=-1$ 
spinning, charged scalarised BHs is bounded by four sets of solutions: 
$i)$ the {\it existence surface}, which corresponds to the set of Kerr-Newman solutions  supporting scalar clouds; 
$ii)$ the set of {\it critical solutions}, which form  again a two dimensional surface;
$iii)$ the {\it static configurations}, $J=0$,
which corresponds to the $\epsilon=-1$ scalarised RN solutions discussed in the previous Section;
and 
$iv)$ the {\it neutral configurations}, $Q=0$,
which were studied with the choice of the coupling function~(\ref{fu}) in~\cite{Berti:2020kgk}.
As for the $Q=0$ case, the {\it  existence surface} is universal for any expression of the coupling function 
allowing for scalarisation.
Concerning the set  $ii)$ ({\it critical solutions}) 
the numerical process fails to converge as it is approached, as in the static  limit.
The explanation for this behaviour 
can again be traced back to the fact that the radicand of a square root
in the horizon expansion of the solutions 
vanishes as the critical set is approached.  
Note also that the sets  
  $ii)$-$iv)$ are not universal; they depend on the choice of the coupling function $f(\phi)$.

\section{Lessons from alternative charged BHs}
\label{sec5}

It is interesting to test the generality
of some of the results above concerning the interplay between the introduction of charge and the scalarisation phenomenon. For this purpose, in this Section we shall be considering the simpler case of static, spherically symmetric, charged BHs in some alternative models, rather than electrovacuum.

As a first observation, we remark that for a finite mass, asymptotically flat solution,
the GB invariant $R^2_{\rm GB}$ is strictly positive for large enough $r$.
As with the RN BH,
a sufficient condition for the occurrence of 
GB induced scalarization for $both$ signs of $\epsilon$
is that $R^2_{\rm GB}<0$ at the horizon.
In general, however, 
the sign of the GB invariant 
 at the horizon  
depends on the matter content.
Indeed, for a generic spherically symmetric BH spacetime
and
using the metric ansatz (\ref{RNn}), 
a straightforward computation leads
to the simple relation
\begin{eqnarray}
R^2_{\rm GB}\big |_{r=r_h}=\frac{12}{r_h^2}+16 \rho_{(H)}^2-\frac{16}{r_h^2}\left[2\rho_{(H)}+p_{\theta (H)}\right] \ ,
\end{eqnarray}
 where $r_h$ is the horizon radius, 
$\rho_{(H)}=-T_t^t(r_h)$
and 
$p_{\theta (H)}=T_\theta^\theta(r_h)$.
One can easily see that, for a generic matter content,
the above quantity has no definite sign.

One may then ask the following two questions: $(1)$ is $R^2_{\rm GB}\big |_{r=r_h}<0$ close to the maximal charge for any charged BH model? $(2)$ is $R^2_{\rm GB}\big |_{r=r_h}<0$ a \textit{necessary} condition for GB$^-$ scalarisation for any charged BH model?

We will now show, by concrete illustrations, that both these questions have a 
$negative$
answer.

\subsection{Einstein-Maxwell-dilaton BHs}

To answer question $(1)$ above, we have investigated the sign of the GB invariant
(together with its behaviour in the bulk) for the 
stringy generalisation of the RN BH -- the Gibbons-Maeda-Garfinkle-Horowitz-Strominger
(GMGHS) family of BHs~\cite{Gibbons:1987ps, Garfinkle:1990qj}.
In our context, these solutions are found for an  action of the form 
\begin{eqnarray}
\label{action2}
\mathcal{S}_{\rm EMdsGB} = \int d^4 x \sqrt{-g} 
\left[
\frac{1}{4}
R -\frac{1}{2}\partial_\mu\psi\partial^\mu\psi -\frac{1}{4}e^{-2\alpha \psi }F_{\mu \nu}F^{\mu \nu}-
 \frac{1}{2}\partial_\mu\phi\partial^\mu\phi 
+\epsilon \frac{\lambda^2 }{4}  f(\phi) R^2_{\rm GB} 
%
\right] \ ,
\end{eqnarray}
describing an Einstein-Maxwell-dilaton-scalar-GB model, 
which includes an extra scalar field (the dilaton $\psi$)
with a non-minimally coupling with the Maxwell term,
where $\alpha\geqslant 0$ is a constant of the theory. 
The GMGHS solution is found for $\phi=0$,  with $f(\phi)$  satisfying the condition (\ref{f-phi-small}).
It is easy to prove that the behaviour found in the RN case ($\alpha=0$, $\psi=0$)
is recovered for small enough values of $\alpha$.
 In that case, for large enough values of the electric charge  $R^2_{\rm GB}$ becomes negative in a region
 between horizon and some maximal value of the radial coordinate.
%
However,
a direct computation shows that, for $\alpha>0.9036$ (for
  the conventions used in~\cite{Garfinkle:1990qj}),
$R^2_{\rm GB}$
is strictly positive at the horizon and also in the bulk,
irrespective of the value of the electric charge.
Thus, as with Schwarzschild vacuum BHs,
 GB$^-$ scalarisation of the 
GMGHS with large enough values 
of the dilaton coupling constant becomes impossible. 
This suggests
 that
 it would be interesting to check the  status of GB$^-$ scalarisation of the rotating counterpart of the GMGHS BH, the well known Kerr-Sen BH~\cite{Sen:1992ua}.

\subsection{Einstein-Yang-Mills BHs}
To answer question $(2)$ above, we have investigated the sign of the GB invariant for the case of Einstein--Yang-Mills  (EYM) 
BHs with
SU(2)
 non-Abelian hair (nA)  \cite{89}.
In our context, these solutions are found for an  action of the form 
\begin{eqnarray}
\label{action2n}
\mathcal{S}_{\rm EYMsGB}= \int d^4 x \sqrt{-g} 
\left[
\frac{1}{4} R -\frac{1}{4} F_{\mu \nu}^{(a)}F^{\mu \nu(a)}-
 \frac{1}{2}\partial_\mu\phi\partial^\mu\phi 
+\epsilon \frac{\lambda^2 }{4}  f(\phi) R^2_{\rm GB} 
%
\right] \ ,
\end{eqnarray} 
with 
$F_{\mu \nu}^{(a)}$
the nA field strength and
$a=1,2,3$.
These so called \textit{coloured} BHs are asymptotically flat and  possess a 
 \textit{single} global ``charge" -- the ADM mass, despite the presence of a local magnetic field
(see \cite{Volkov:1998cc,Volkov:2016ehx} 
for reviews).  
Another striking difference with respect to their (magnetic) 
RN Abelian counterparts is 
the existence of a smooth solitonic limit  \cite{Bartnik:1988am}, obtained
as the horizon size shrinks to zero.
At the same time, there is no upper bound on their horizon size.
However, the large  EYM BHs  are essentially Schwarzschild solutions;
the contribution of the YM fields to the total ADM mass becomes negligible,  albeit these fields
are still nontrivial.
 
\begin{figure}[t!]
\hbox to\linewidth{\hss%
	\resizebox{9cm}{7cm}{\includegraphics{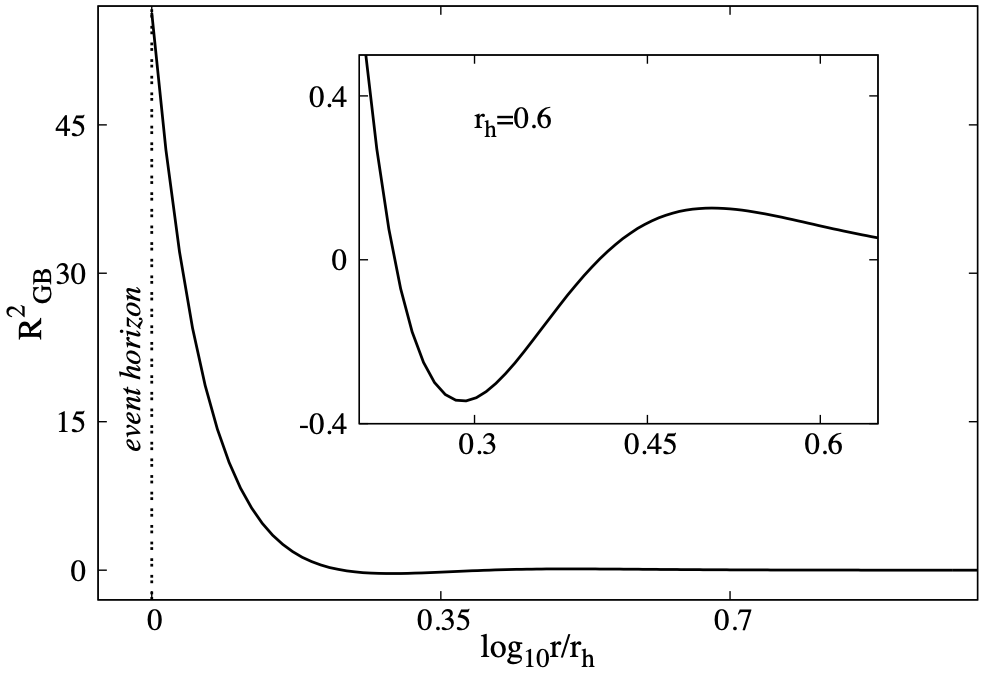}} 
 	\resizebox{9cm}{7cm}{\includegraphics{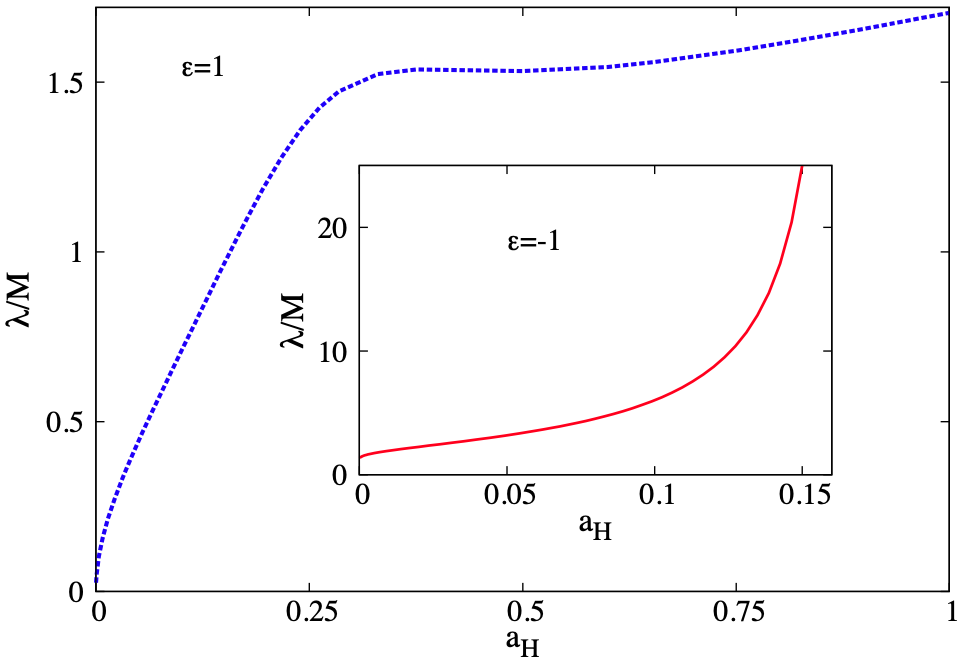}}  
\hss}
\caption{\small 
{\it Left panel:}  GB invariant
 as a function of 
the radial coordinate for an  EYM  BH
with horizon radius $r_h=0.6$. 
One notices a shell with $R^2_{GB}<0$, located outside the horizon.
{\it Right panel:}  $\lambda/M$ of the set of EYM  BHs
supporting scalar clouds 
as a function of the reduced horizon area for $\epsilon=\pm 1$.
}
\label{nA}
\end{figure}

Contact with question $(2)$ above comes from observing that  
the
  GB invariant 
is always positive at the horizon for these solutions.
However, $R^2_{\rm GB}$ may take negative values 
	in a shell which does not touch the horizon, $i.e.$ for some range of the radial coordinate 
$r_h<r_1<r<r_2$ -  see Fig. \ref{nA} (left panel).
This feature occurs for small enough BHs: using the metric form  (\ref{RNn})
and the conventions in \cite{Volkov:1998cc}, we confirmed such shell is present for 
$0<r_h\leqslant 0.71$,
or, equivalently,
$0<a_H\leqslant 0.158$.

Given this qualitative difference with the RN case, one could ask whether GB$^\epsilon$ scalarisation is still possible  for both signs of $\epsilon$.  
The answer is positive, and 
we have constructed the corresponding  GB scalar clouds, $i.e.$ solved eq. (\ref{eq-phi-small})
for a large set of EYM BH backgrounds 
and both values of $\epsilon$ - see Fig. \ref{nA} (right panel).
As expected,  $\epsilon=+1$
scalar clouds exist for  all non-Abelian BHs. 
The value of the ratio $\lambda/M\simeq 1.704$
corresponding to a Schwarzschild BH
is approach asymptotically, as  $a_H\to 1$ ($i.e.$ large EYM BHs).
Also, scalar clouds with  
$\epsilon=-1$ 
exist for {\it all} BHs with  $R^2_{\rm GB}<0$ in a shell outside the horizon.
Nonlinear continuations of these scalar clouds should exist, 
but we did not attempt to construct them.

This example makes clear that GB$^\epsilon$ scalarisation of BHs is not necessarily supported and triggered in the immediate vicinity of the event horizon.

 {\small \hspace*{3.cm}{\it  } }

\section{Further remarks}
\label{sec6}

There is a well known analogy between the 
spinning vacuum Kerr BHs and the  electrically charged static RN solutions.\footnote{Since 
the electric-magnetic duality is still valid for the (Abelian) models in this work,
the solutions possess a dual magnetic description.}
They possess many similar properties at the level of a thermodynamical description;
in particular, both RN and Kerr BHs have an extremal limit, with a finite horizon size.
In the context of this work,
it is interesting to note that the GB invariant of a RN BH changes sign 
if $q$ is 
 large enough as the Kerr one changes sign for sufficiently large $j$.
Thus, 
 it is natural to 
conjecture that the qualitative picture found concerning the GB$^\epsilon$ scalarisation of Kerr BHs~\cite{Cunha:2019dwb,Dima:2020yac,Herdeiro:2020wei,Berti:2020kgk}
should be essentially recovered when  replacing rotation by electric charge,
with the existence of both $\epsilon=\pm 1$ scalarised  solutions.

 In this work we have confirmed that this conjecture is true,
and
construct the corresponding scalarised 
RN BHs.
That is,
we provide evidence for the following scenario:
given an expression for the coupling function
$f(\phi)$,
two classes of charged RN-scalarised solutions
may exist for the same global charges.
The first one has 
 $
\epsilon= 1,
$
and can be viewed as
a generalization of the $Q=0$ solutions in 
 \cite{Silva:2017uqg,Doneva:2017bvd,Antoniou:2017acq}.
The second has
 $
\epsilon=- 1,
$
and in this case 
the condition $\mu_{\rm eff}^2<0$
is supported by a large enough charge to mass ratio $q>q_c=0.957058$,
which implies $R^2_{\rm GB}<0$
for some region outside the horizon.
We have also presented a
 preliminary
 investigation of 
the spinning generalisations of the above BHs, $i.e.$, the scalarised 
Kerr-Newman BHs.

In the last part of this work,
we have addressed the generality of these results.
First, we established that the GB invariant of the stringy
dilatonic generalisation of the RN BH becomes strictly positive for 
large enough values of the dilaton coupling constant. 
Also, we pointed out the possibility that the $\epsilon=-1$
 BH scalarisation may also appear 
in situations where $R^2_{\rm GB}$
is negative in a spherical $shell$ outside and disconnected from the horizon.
This is the case of the coloured BHs in EYM theory.

In this paper, to simplify the picture, we have assumed 
the absence of a self-interaction term for the scalar field in the action (\ref{action}).
GB$^+$ scalarisation of spherical BHs including such self-interactions 
is discussed $e.g.$ in   \cite{Doneva:2019vuh,Macedo:2019sem}. Our results could be generalised to include such
self-interactions.

\section*{Acknowledgements}

This work is supported  by the  Center for Research and Development in Mathematics and Applications (CIDMA) through the Portuguese Foundation for Science and Technology (FCT -- Fundac\~ao para a Ci\^encia e a Tecnologia), references  UIDB/04106/2020 and UIDP/04106/2020,
 and by national funds (OE), through FCT, I.P., 
in the scope of the framework contract foreseen in the numbers 4, 5 and 6 of the article 23,of the Decree-Law 57/2016, of August 29, changed by Law 57/2017, of July 19.  
The authors acknowledge support  from the projects PTDC/FIS-OUT/28407/2017, CERN/FIS-PAR/0027/2019 and PTDC/FIS-AST/3041/2020. A. Pombo is supported by the
FCT grant PD/BD/142842/2018.  This work has further been supported by  the  European  Union's  Horizon  2020  research  and  innovation  (RISE) programme H2020-MSCA-RISE-2017 Grant No.~FunFiCO-777740. The authors would like to acknowledge networking support by the COST Action CA16104.


 \begin{small}
 
 \end{small}

 \end{document}